\documentclass[prl,twocolumn,showpacs,byrevtex,amsmath,amssymb,nofootinbib]{revtex4}

\usepackage{graphicx}
\usepackage{dcolumn}
\usepackage{bm}

\begin{document}

\title{Lane formation in driven colloidal mixtures: is it continuous or discontinuous?}

\author{Martin Rex$^1$, C. Patrick Royall$^2$, Alfons van Blaaderen$^3$, Hartmut L\"owen$^1$}

\affiliation{ $^1$Institut f\"ur Theoretische Physik II: Weiche Materie, Heinrich-Heine-Universit\"at
  D\"usseldorf, Universit\"atsstrasse 1, D-40225 D\"usseldorf, Germany \\
  $^2$School of Chemistry, University of Bristol, Bristol, BS8 1TS, UK \\
  $^3$Soft Condensed Matter Group, Debye Institute for Nanomaterials Science,
  Utrecht University, PO Box 80000, 3508 TA Utrecht, The Netherlands
}

\date{\today}

\begin{abstract}

Binary mixtures of oppositely charged colloids 
driven by an electric field are shown to exhibit a 
nonequilibrium transition towards
lane formation if the driving force is increased.
Brownian dynamics computer simulations and real-space experiments are employed
to study hysteresis effects in an 
order parameter measuring the extent of lane formation upon increasing and decreasing
the driving force. Both from  simulation and experiment,
we find that lane formation due to electrical fields is continuous. However, simulations 
show a discontinuous transition if the driving force is gravity. 
\end{abstract}


\maketitle


In equilibrium, there is a fundamental difference between a sharp phase transition which
exhibits a jump in a certain derivative of the free energy with respect to a thermodynamic variable
\cite{thermodynamics_books}  and a continuous cross-over where no such discontinuity
exists in the thermodynamic limit.
This is much less clear for nonequilibrium phase transitions since a free 
energy does not exist in general in nonequilibrium. Here, in many situations,
 an order parameter can still be defined in the steady state
and hysteresis behaviour typically serves as a criterion to discriminate between 
discontinuous and continuous behaviour  \cite{Zia}.
While it is by now well-understood how the order of an equilibrium
 phase transition is controllable by the interparticle interactions \cite{equi1,equi2}, 
the question of which key parameters determine the existence and order of 
nonequilibrium phase transitions is much less understood 
\textbf{due to the multitude} of parameters 
characterizing the \textbf{nature} of the dynamics which do not apply in equilibrium.

Lane formation in a binary mixture of Brownian particles 
which are driven by a constant external force depending on the particle species
 \cite{dzubiella_pre_2002} represents a prototype 
of a nonequilibrium transition in a continuous (i.e.\ off-lattice) system \cite{Chaikin_Pine}. 
Brownian dynamics computer simulations strongly
support the scenario that for increasing driving force
 the system undergoes a transition from a mixed steady state
 towards a steady state where macroscopic lanes are formed.
Such a laning  transition has been found in models for oppositely driven repulsive mixtures
\cite{dzubiella_pre_2002,netz_epl_2003,dzubiella_epl_2003,dzubiella_fd_2003,
delhommelle_pre_2005,pandley_ijmpc_2003} in two and three spatial dimensions.
 In two spatial dimensions, it was found  \cite{dzubiella_pre_2002} that a suitable order parameter 
which detects laning exhibits a significant hysteresis - if the driving force is 
increased and subsequently decreased - which signals a discontinuous 
nonequilibrium phase transition. This issue is still unclear in three spatial dimensions.
Though the general
scenario  occurs
for  pedestrian dynamics \cite{helbing_pa_2006,helbing_njp_2003},
 in driven granular  \cite{ehrhardt_pre_2005,coniglio_prl_2005} and colloidal
 \cite{leunissen_nature_2005} matter and in 
complex plasmas \cite{Morfill_Buch}, a quantitative comparison of laning
with simulation data and the determination of the underlying order of the transition
is still lacking.

The aim of this letter is twofold: first, we show that oppositely charged colloidal 
particles driven in an electric field \cite{leunissen_nature_2005}
exhibit  lane formation in quantitative agreement with our Brownian dynamics computer simulations.
Second, we address the question whether the transition towards lane formation is discontinuous or 
a smooth crossover. While charged colloids  are found to exhibit lanes in a 
continuous way upon increasing the electric field strength, 
additional simulations reveal that the transition becomes discontinuous
if gravity is the driving force. The physical reason is that hydrodynamic 
interactions are screened in the electric context \cite{long_epje_2001,rex_EPJE_2008} while 
they are long-ranged for gravity. In fact, three-dimensional simulations which 
neglect hydrodynamic interactions completely yield a continuous transition as well.
Our results demonstrate that the existence and order of
a nonequilibrium phase transition  depends on details of the dynamics even when the
particle interactions which entirely determine the order of phase transitions in equilibrium 
are kept fixed.

In our experiments, we used suspensions of polymethylmethacrylate colloids in a density matching mixture 
of cyclohexyl bromide and cis decalin into which we dissolved 60 $\mu$M tetra butyl ammonium bromide salt. 
An equimolar binary suspension of colloids fluorescently 
labelled with Rhodamine and 7-nitrobenzo-2-oxa-1,3-diazol with volume fraction $\eta_N=\eta_R=0.1$ 
was prepared. The diameter of both species was 1.2 $\mu$m. The inclusion of salt led to 
both colloidal species accquiring opposite charges of $Z_i = \rm 100 e$, and a Debye screening length
of $\kappa^{-1} \approx 300$ nm \cite{leunissen_nature_2005}.
The solvent dielectric constant $\epsilon$ was $5.5$, and shear viscosity was $\eta=2.2$ cP. All experiments were
conducted at room temperature $T=300$ K. We used a Leica NT confocal laser scanning microscope. Experimental data
was taken in 2D scans of typically 100 $\mu$m in width at a frame rate of around 2 \textbf{frames per second},
ie coordinates were collected from a diffraction-limited slice of $\sim 2 \mu m$ in depth.

In our Brownian dynamics computer simulations, we consider
an equimolar binary mixture of $N=1024$ oppositely charged colloidal particles of
hard-core diameter $\sigma$  at corresponding total volume fraction $\phi=0.2$
dispersed in a solvent whose viscosity and dielectric constant
matched the experimental values, likewise we set and $T=300$ K.

The simulation is performed in a cubic box of length $l$ with 
periodic boundary conditions in all three directions \cite{rex_pre_2007}.
Apart from their steric repulsion, the particles interact via the 
screened Coulomb potential
\begin{equation}
  \label{eq:yukawa_potential}
  V_{ij}(r) =\frac{Z_{i}Z_{j}}{\epsilon (1+\kappa\sigma/2)^2}
  \frac{e^{(-\kappa\sigma(r/\sigma-1))}}{r}
\end{equation}
where $i$ and $j$ label the particle species, $Z_{i}$ denotes the particle charge,
$r$ is the interparticle separation and $\kappa$ is the inverse 
Debye-H\"uckel screening length depending on the salt concentration in the solvent.
We take $\kappa \sigma =4$ in order to match the experimental salt concentration.
An electric field $\vec E$ yields the driving forces 
$\mathbf{F}_{i}^{\mathrm{ext}} = Z^*_i{\vec E}$ acting on the particles.
The actual
electrophoretic charge renormalization is taken to be $Z^*_i / Z_i = 0.8$
as obtained from the averaged drift velocity in the electric field at low densities
\cite{leunissen_thesis}. 

Time-dependent trajectories of the particles are calculated 
using a finite-time step method with a
configuration-dependent diffusion tensor \cite{allen_tildesley_book}.
The magnitude of friction is set by the Stokes-drag expression where 
the free diffusion constant is given by $D_0 k_BT/3\pi\eta\sigma_H$ 
(with $k_B$ denoting Boltzmann's constant) which sets the typical Brownian time scale 
$\tau_B = D_0/\sigma^2$. Here the hydrodynamic diameter of the particles is set very close to the 
interaction core $\sigma$ such that $\sigma_H = 0.98 \sigma$.
We use the pairwise approximation based  the Long-Ajdari mobility
tensor \cite{long_epje_2001} in order to include hydrodynamic interactions properly 
for a driving electric field
\cite{rex_EPJE_2008}. After an initial relaxation period of typically
$20\tau_{\mathrm{B}}$  the system runs into a 
steady state. The time-step used in the simulation was $\Delta t = 10^{-4} \tau_B$.

Experimental  snapshots in the steady state  for two electric field strengths
$E = 30 kV/m$ and $E = 100 kV/m$
are presented in Figure 1. The vertical extent of the optical slice is 
set by diffraction, leading to a lengthscale of order $2\mu m$.
The snapshots clearly reveal the onset of laning for increasing electric driving field.
Corresponding snapshots of our Brownian dynamics computer simulations are presented in Figure 2.
For sake of comparison, the same field strengths were used as for the experimental
snapshots.
The simulation configurations
show a similar tendency towards laning along the drive direction 
when the electric field is increased.

\begin{figure}[h!]
  \begin{center}
    \includegraphics[width=7cm]{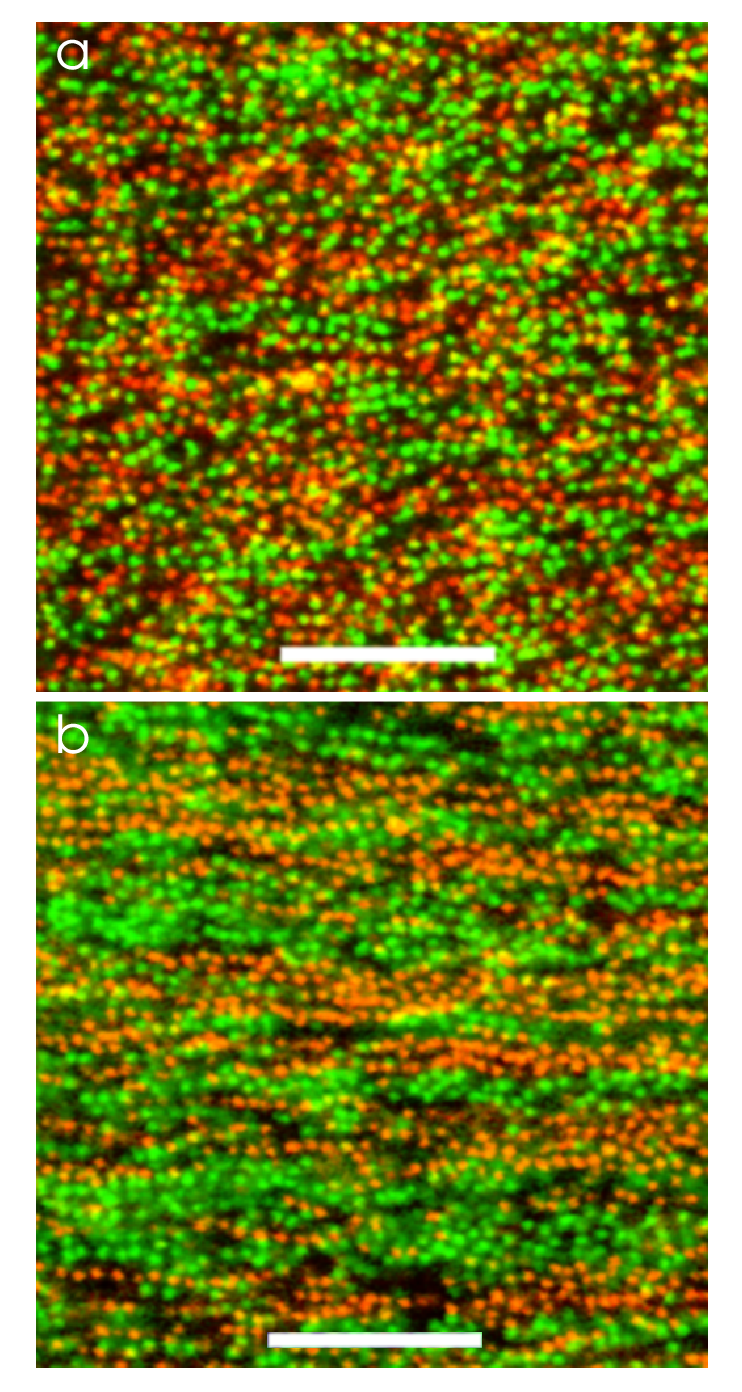}
    \caption{Typical particle configuration for positively (green spheres) and negatively (red spheres)
      charged colloids in an electric field of strength $E$ along the horizontal direction.
      Only particles whose centres are within a slice of thickness $2 \mu m$ are shown. 
      The length bar is 10 $\mu m$. The formation
      of lanes can clearly be seen.  (a) experimental snapshot for  $E = 30 kV/m$, 
      (b) experimental 
      snapshot for  $E = 100 kV/m$. The other parameters are given in the text.}
    \label{}
  \end{center}
\end{figure}

\begin{figure}[h!]
  \begin{center}
    \includegraphics[width=7cm]{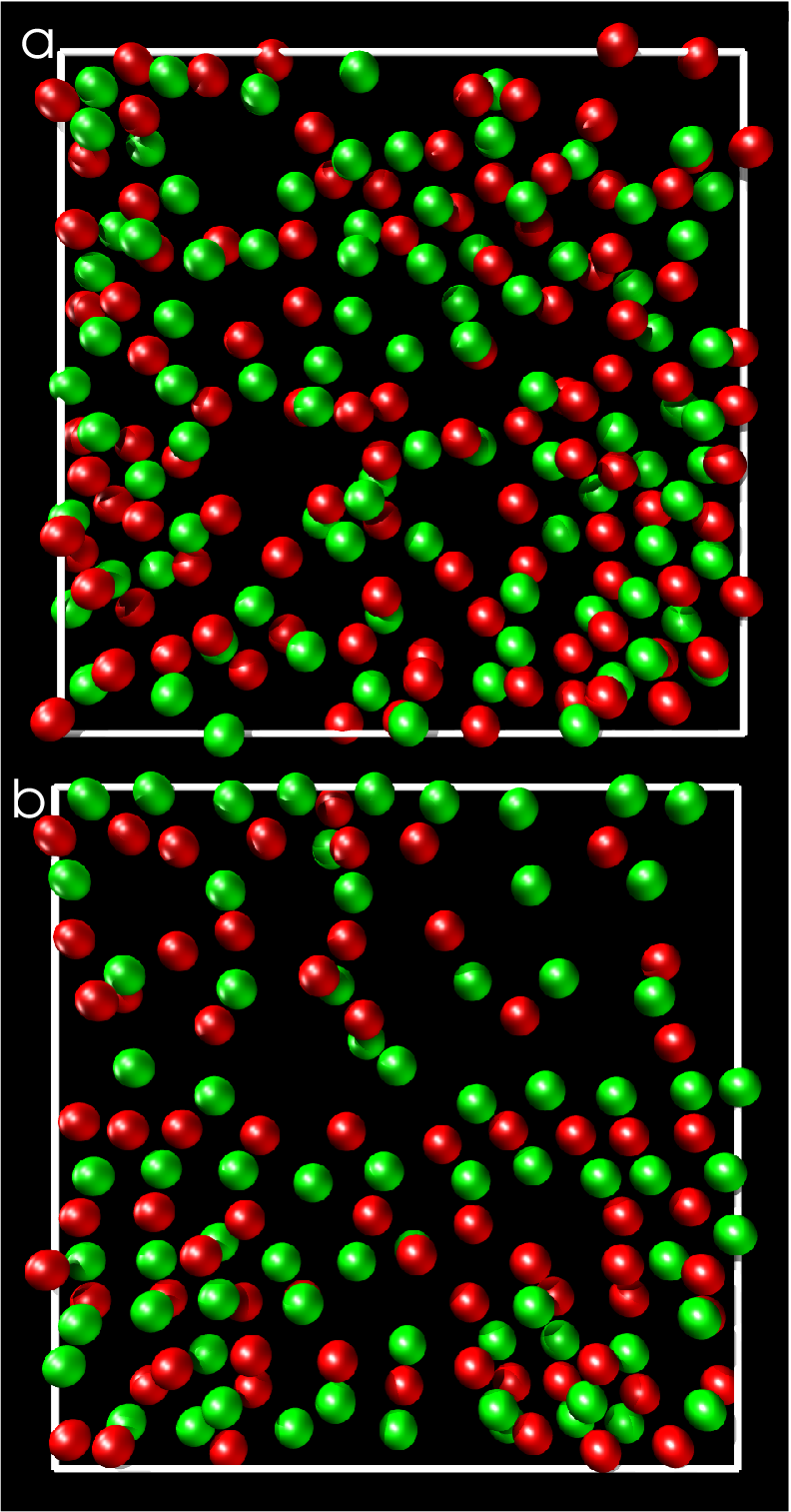}
    \caption{Computer simulation snapshots for positively (green spheres) and negatively (red spheres)
      charged colloids in an electric field of strength $E$ along the horizontal direction.
      Only particles whose centres are within a slice of thickness $2\sigma$ are shown.
  (a)   $E = 30 kV/m$, 
      (b)   $E = 100 kV/m$.  The other parameters are given in the text.}
    \label{}
  \end{center}
\end{figure}

In order to quantify the extent of laning,
we define a 
laning order parameter $m$ as follows: we assign a cylinder of diameter $0.9\sigma$
and total height $6\sigma$ symmetrically 
around the center of each particle $i$ such that the cylinder axis is parallel to the driving field.
If one or more particles of different species are contained in this cylinders, we define
the value of $m_i=0$ to this particle while $m_i=1$ if the cylinder is free from other particles
with an opposite charge. The averaged dimensionless laning order parameter 
\begin{equation}
  \label{eq:order_parameter}
m = <\sum_{i=1}^N m_i>/N
\end{equation}
 where $<...>$ refers 
to a particle and steady-state average and $N$ is the total number of particles considered
 measures the averaged extent of laning. By construction,
the laning order parameter $m$  equals 1 if particles of the same 
species are always on top of each other along the drive but it vanishes in a completely mixed 
situation. We have 
averaged this laning order parameter  over many simulation
configurations. For the experimental data, a similar  laning order parameter was obtained
by using a projected cylinder (i.e. a rectangular cut) to the plane containing the particles. 
Plots of the laning order parameter $m$ versus the strengths of the electric 
field applied are presented in Figure 3.

\begin{figure}[h!]
  \begin{center}
    \includegraphics[width=8cm, clip=true, draft=false]{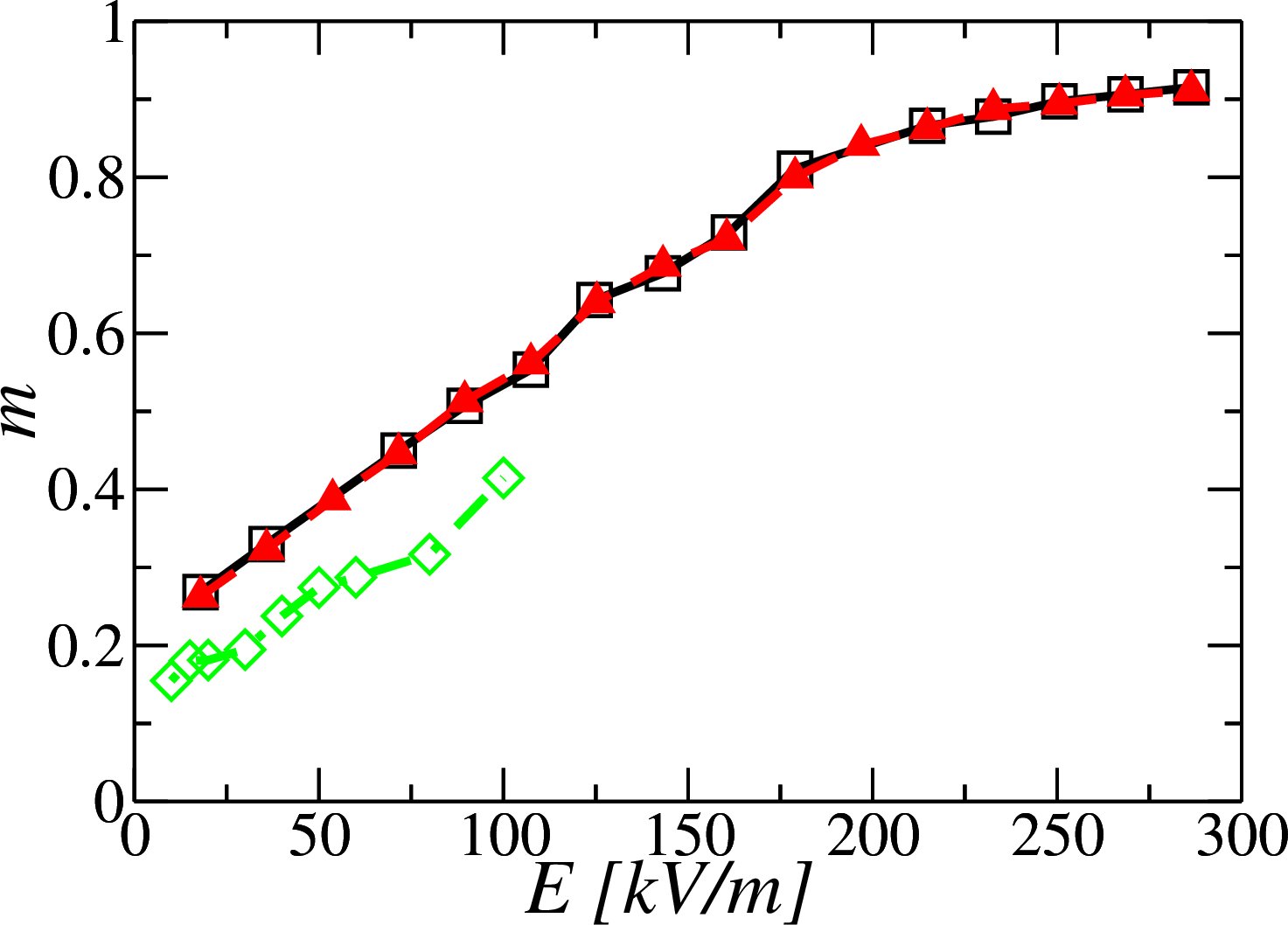}
    \caption{Averaged laning order parameter $m$ versus electric field strength
      $E$ (in units of $kV/m$). The open squares (full line) are the simulation results upon 
      increasing the field strength, the full triangles (dashed line) 
      shows the same  upon decreasing the field strength. No hysteresis is found.
      The diamonds (dot-dashed line)  are the experimental data. 
      The parameters are the same as in Fig.\ 1.}
    \label{}
  \end{center}
\end{figure}


First, there is reasonable agreement between simulation and experiment. 
The increase of the laning order parameter with the external drive is smooth.
The experimental data reveal a slightly smaller laning order parameter
than the simulation which is acceptable given the uncertaincy in the
particle charge. The experimental data were
taken both increasing and decreasing the external field, and no difference
was observed. Hence we conclude that effects of hysteresis are largely
absent. We remark that possible effects arising from electro-osmotic
flow are not exactly known. 


\begin{figure}[h!]
  \begin{center}
    \includegraphics[width=8cm, clip=true, draft=false]{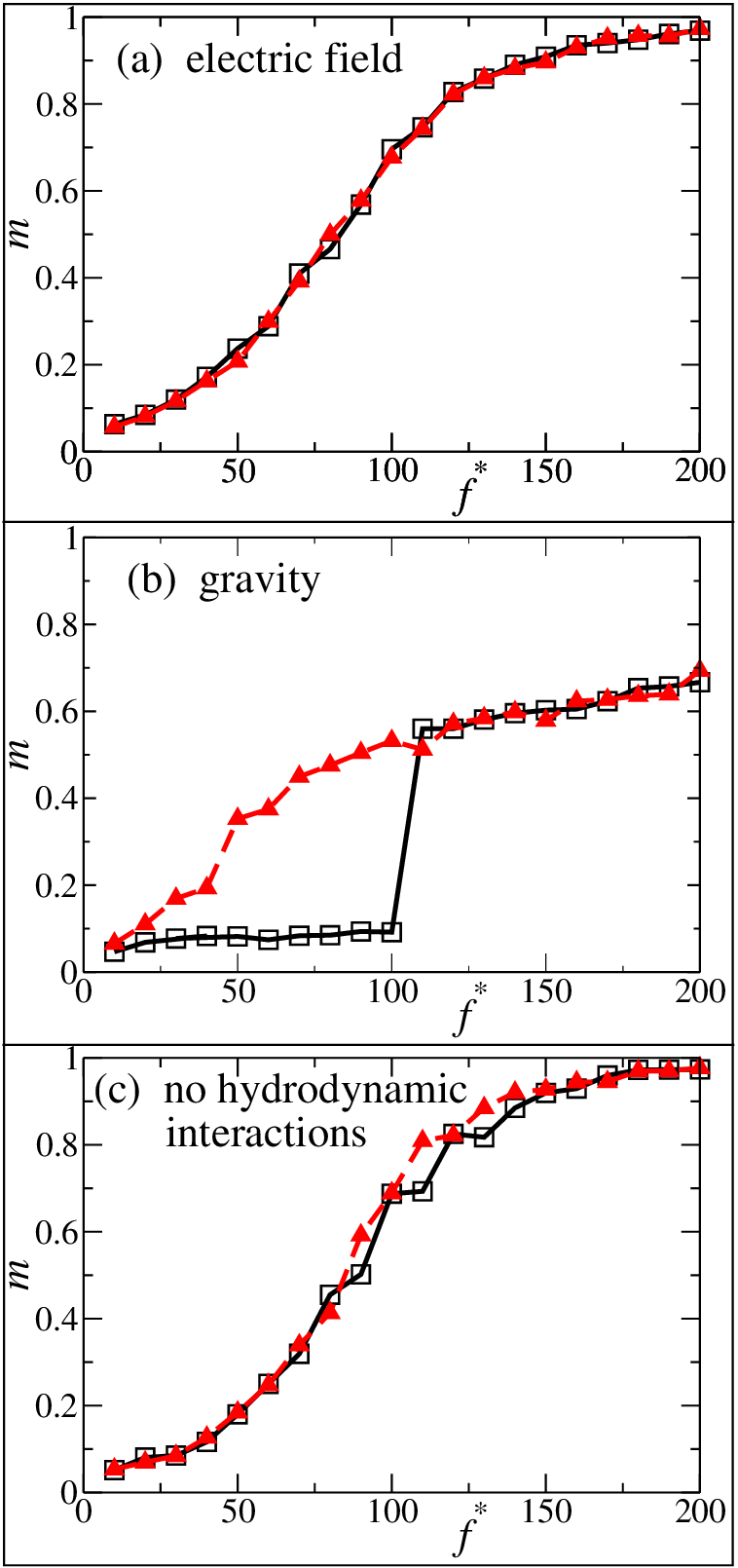}
    \caption{Averaged laning order parameter versus reduced driving force $f^*$ 
      both upon increasing (open squares, full line) and decreasing (full triangles, dashed line) 
      the driving forces: (a) for an electric driving field where
      $f^*= E \mod Z^*_1 \mod \sigma / k_B T$ and hydrodynamic interactions are screened, 
      (b) for a gravitational force $M g$ where $f^*=\pm Mg\sigma /k_B T$ and hydrodynamic interactions 
      are long-ranged, (c) same as (a)/(b) but for neglected hydrodynamic interactions.
      The length of the cylinder in the definition of the order parameter here spans the whole
      simulation box.}
  \end{center}
\end{figure}

Finally, we consider the case of gravity in which the driving forces
result from a different buoyant mass of the two particle species. In particular, we treat
the symmetric case here where plus and minus charged particles feel the same gravitational 
force but just their sign is different.
Our Brownian dynamics computer simulations are now performed with 
the unscreened Rotne-Prager mobility tensor \cite{rotne_jcp_1969} 
replacing the screened Long-Ajdari tensor. All the other parameters are kept unchanged.
In Figure 4 the simulation results for the laning order parameter are shown
for increasing and decreasing driving forces. The results are compared 
to the case of an electric field where the mobility tensor is  screened
due to counterion counter-motion.
A significant hysteresis behaviour is detected for sedimentation
which clearly signals a discontinuous transition towards laning while no
such hysteresis is present for electrophoresis and in the complete 
absence of hydrodynamic interactions. This clearly indicates that it is the
range of hydrodynamic interactions which make lane formation discontinuous.
In a strictly two dimensional system, laning
occurs in a discontinuous way even when hydrodynamic interactions are ignored completely
 \cite{dzubiella_pre_2002}. Therefore besides the hydrodynamics the system dimensionality
plays an important role in determining whether laning is discontinuous or continuous.

In conclusion, we have shown that oppositely charged colloids driven by an external electric field
show a continuous tendency towards lane formation if the driving strength is increased.
Real-space experimental data are in quantitative agreement with Brownian dynamics 
computer simulations which include the screened hydrodynamic interactions between the driven 
colloids. The continuous crossover towards laning is in marked contrast to sedimenting colloids
where lane formation occurs via a nonequilibrium first-order phase transition
 as revealed by a significant hysteresis
behaviour in a suitable order parameter. This change must be attributed to the long-range 
nature of hydrodynamic interactions in the sedimentation case. In a reduced system dimensionality, 
on the other hand, laning is discontinuous as well even when hydrodynamic interactions are ignored.
Hence the order of the lane formation is determined by both, hydrodynamic interactions and system 
dimensionality. Therefore the more general conclusion is that the existence and order of
a nonequilibrium phase transition  depends on details of the dynamics even when the
particle interactions which entirely determine the order of phase transitions in equilibrium 
are kept fixed.

Lane formation does not only occur in colloidal systems but also in
dusty plasmas where hydrodynamic interactions are absent but inertia effects play a dominant role
\cite{Morfill_Buch}. We expect that, in three dimensions, the tendency towards lane formation will
be continuous there as well. In strict two dimensions, e.g. for pedestrian dynamics 
and shaken granular matter, we expect a continuous behaviour irrespective to the 
details of the dynamics.

\acknowledgements

We thank A. Ladd and A. Wysocki for helpful discussions. This work was been supported by the SFB TR6
(DFG/FOM) within the projects A3 and D1. 

\bibliographystyle{prsty}

\end{document}